\documentstyle[a4,12pt,epsfig]{article}
\begin{document}


\vspace{0.2cm}

\begin{center}
{\Large\bf Neutrino Mixing: from the Broken $\mu$-$\tau$ Symmetry
to the Broken Friedberg-Lee Symmetry} \footnote{Talk given at the
International Workshop on Neutrino Masses and Mixings, December 17
-- 19, 2006, Shizuoka, Japan}
\end{center}

\vspace{0.5cm}

\begin{center}
{\bf Zhi-zhong Xing} \footnote{
E-mail: xingzz@ihep.ac.cn}\\
{\sl Institute of High Energy Physics, Chinese Academy of
Sciences, \\
P.O. Box 918, Beijing 100049, China}
\end{center}

\vspace{1.5cm}

\begin{abstract}
I argue that the observed flavor structures of leptons and quarks
might imply the existence of certain flavor symmetries. The latter
should be a good starting point to build realistic models towards
deeper understanding of the fermion mass spectra and flavor mixing
patterns. The $\mu$-$\tau$ permutation symmetry serves for such an
example to interpret the almost maximal atmospheric neutrino
mixing angle ($\theta^{}_{23} \sim 45^\circ$) and the strongly
suppressed CHOOZ neutrino mixing angle ($\theta^{}_{13} <
10^\circ$). In this talk I like to highlight a new kind of flavor
symmetry, the Friedberg-Lee symmetry, for the effective Majorana
neutrino mass operator. Luo and I have shown that this symmetry
can be broken in an oblique way, such that the lightest neutrino
remains massless but an experimentally-favored neutrino mixing
pattern is achievable. We get a novel prediction for
$\theta^{}_{13}$ in the CP-conserving case: $\sin\theta^{}_{13} =
\tan\theta^{}_{12} |(1- \tan\theta^{}_{23})/ (1+
\tan\theta^{}_{23})|$. Our scenario can simply be generalized to
accommodate CP violation and be combined with the seesaw
mechanism. Finally I stress the importance of probing possible
effects of $\mu$-$\tau$ symmetry breaking either in terrestrial
neutrino oscillation experiments or with ultrahigh-energy cosmic
neutrino telescopes.
\end{abstract}

\newpage

\section{``Who ordered that?"}

There are five categories of fundamental particles that we have
known today: (a) charged leptons $e$, $\mu$ and $\tau$; (b)
neutrinos $\nu^{}_e$, $\nu^{}_\mu$ and $\nu^{}_\tau$; (c) up-type
quarks $u$, $c$ and $t$; (d) down-type quarks $d$, $s$ and $b$;
(e) gauge bosons $\gamma$, $W^\pm$, $Z^0$ and $g$. The particles
belonging to category (e) are the carriers of electromagnetic
($\gamma$), weak ($W^\pm$ and $Z^0$) and strong ($g$) forces,
respectively. In comparison, the particles belonging to categories
(a) -- (d) are the building blocks of the world of matter. But we
have no idea about what constitutes the cold dark matter of the
Universe.

The afore-mentioned fermions have three families: the first family
includes $u$, $d$, $e$ and $\nu^{}_e$; the second family contains
$c$, $s$, $\mu$ and $\nu^{}_\mu$; and the third family is composed
of $t$, $b$, $\tau$ and $\nu^{}_\tau$. The leptons or quarks of
each category have the same gauge quantum numbers, but their
masses are different from one another. What distinguishes
different fermion families? In other words, ``Who ordered
that?"\footnote{Such a question was first asked by the famous
American physicist I.I. Rabi, when he heard that the muon ($\mu$),
a sister of the electron, was discovered from cosmic rays in
1936.} Although the definite answer to this fundamental question
has been lacking, some people have speculated the possibility that
there might exist certain hidden flavor quantum numbers or flavor
symmetries which distinguish one family from another.

The mysterious Koide relation for the pole masses of three charged
leptons can be taken as a good example to illustrate the puzzles
of flavors \cite{Koide1}:
\begin{eqnarray}
Q^{\rm pole}_l \; \equiv \; \frac{m^{}_e + m^{}_\mu + m^{}_\tau}{
\left (\sqrt{m^{}_e} + \sqrt{m^{}_\mu} + \sqrt{m^{}_\tau} \right
)^2} \; = \; \frac{2}{3} \;\; .
\end{eqnarray}
Given the experimental values \cite{PDG}
\begin{eqnarray}
m^{}_e & = & \left (0.510998918 \pm 0.000000044 \right ) ~ {\rm
MeV} \; , ~~
\nonumber \\
m^{}_\mu & = & \left (105.6583692 \pm 0.0000094 \right ) ~ {\rm
MeV} \; ,
\nonumber \\
m^{}_\tau & = & \left (1776.99^{+0.29}_{-0.26} \right ) ~ {\rm
MeV} \; ,
\end{eqnarray}
Eq. (1) holds up to the accuracy of ${\cal O}(10^{-5})$; i.e.,
$-0.00001 \leq Q^{\rm pole}_l - 2/3 \leq +0.00002$ \cite{Zhang}.
This precision is so amazing that I cannot help to ask what the
underlying physics is behind the Koide relation. As pointed out by
Koide himself in this workshop \cite{Koide2}, Eq. (1) might
naturally stem from a kind of flavor symmetry, such as the
discrete S(3) symmetry. I stress that one has to distinguish
between the concepts of running and pole masses of fermions when
building models at certain energy scales, where certain flavor
symmetries exactly exist.

Needless to say, flavor symmetries may serve as a promising
guiding principle for model building and their spontaneous or
explicit breaking schemes may help understand the observed mass
spectra and flavor mixing patterns of leptons and quarks
\cite{Review}. Here let me focus on neutrino mixing, which is
described by the $3\times 3$ Maki-Nakagawa-Sakata (MNS) matrix
\cite{MNS} at low energy scales. In September 1996, Fritzsch and I
proposed the so-called ``democratic" neutrino mixing pattern by
starting from the $\rm S(3)^{}_{\rm L} \times S(3)^{}_{\rm R}$
symmetry of the charged-lepton mass matrix and the S(3) symmetry
of the neutrino mass matrix \cite{FX96}:
\begin{equation}
V^{}_{\rm FX} \; =\; \left ( \matrix{ \frac{1}{\sqrt{2}} & ~
\frac{1}{\sqrt{2}} ~ & 0 \cr \frac{-1}{\sqrt{6}} &
\frac{1}{\sqrt{6}} & \frac{1}{\sqrt{6}} \cr \frac{1}{\sqrt{3}} &
\frac{-1}{\sqrt{3}} & \frac{1}{\sqrt{3}} \cr } \right ) \; .
\end{equation}
A few years later, Harrison, Perkins and Scott \cite{TB} advocated
the so-called ``tri-bimaximal" neutrino mixing pattern,
\begin{equation}
V^{}_{\rm HPS} \; =\; \left ( \matrix{ \frac{2}{\sqrt{6}} & ~
\frac{1}{\sqrt{3}} ~ & 0 \cr \frac{-1}{\sqrt{6}} &
\frac{1}{\sqrt{3}} & \frac{1}{\sqrt{2}} \cr \frac{1}{\sqrt{6}} &
\frac{-1}{\sqrt{3}} & \frac{1}{\sqrt{2}} \cr } \right ) \; ,
\end{equation}
which can be derived from certain discrete flavor symmetries (such
as the non-Abelian $\rm A^{}_4$ symmetry \cite{A4}). In comparison
with $V^{}_{\rm FX}$, $V^{}_{\rm HPS}$ is much more compatible
with today's solar \cite{SNO}, atmospheric \cite{SK}, reactor
\cite{KM} and accelerator \cite{K2K} neutrino oscillation
data.\footnote{Radiative corrections to both $V^{}_{\rm FX}$ and
$V^{}_{\rm HPS}$ have been analyzed by Luo, Mei and me \cite{Mei}
from a superhigh energy scale to the electroweak scale.} In
particular, the large-mixing-angle Mikheyev-Smirnov-Wolfenstein
(MSW) solution \cite{MSW} to the solar neutrino problem can simply
be accommodated and the very special values of two mixing angles
$\theta^{}_{13} = 0^\circ$ and $\theta^{}_{23} = 45^\circ$ are
closely related to the $\mu$-$\tau$ permutation symmetry.

\section{$\mu$-$\tau$ permutation symmetry}

In the limit of $\mu$-$\tau$ permutation symmetry, the (effective)
Majorana neutrino mass matrix $M^{}_\nu$ takes the form
\begin{equation}
M^{}_\nu \; = \; \left ( \matrix{ A & B & B \cr B & C & D \cr B &
D & C \cr } \right ) \; .
\end{equation}
Here and hereafter we work in the basis where the flavor
eigenstates of charged leptons are identified with their mass
eigenstates. It is easy to check that $M^{}_\nu$ is invariant
under the $\mu$-$\tau$ permutation, which is equivalent to the
transformation $O^{}_\nu M^{}_\nu O^T_\nu$ with
\begin{equation}
O^{}_\nu \; =\; \left ( \matrix{ 1 & ~0~ & 0 \cr 0 & 0 & 1 \cr 0 &
1& 0 \cr } \right ) \; .
\end{equation}
After a straightforward diagonalization of $M^{}_\nu$, one may
obtain $\theta^{}_{13} = 0^\circ$ and $\theta^{}_{23} = 45^\circ$
for the MNS matrix $V$. The solar neutrino mixing angle
$\theta^{}_{12}$ is not fixed, nevertheless.

The breaking of $\mu$-$\tau$ symmetry will in general lead to
$\theta^{}_{13} \neq 0^\circ$, $\theta^{}_{23} \neq 45^\circ$ and
a non-trivial CP-violating phase $\delta$ \cite{MT,XZZ,Zhou,Luo}.
If the $\mu$-$\tau$ symmetry is softly broken, however,
$\theta^{}_{13} = 0^\circ$ or $\theta^{}_{23} = 45^\circ$ might
survive. For instance, it is possible to break the $\mu$-$\tau$
symmetry in a proper way such that the resultant MNS matrix takes
one of the following two forms \cite{XZZ,Zhou,Luo,Baba}:
\begin{equation}
V \; = \; \left ( \matrix{ c^{}_{12} & s^{}_{12} & 0 \cr
-s^{}_{12}/\sqrt{2} & c^{}_{12}/\sqrt{2} & 1/\sqrt{2} \cr
s^{}_{12}/\sqrt{2} & ~~ -c^{}_{12}/\sqrt{2} ~~~ & 1/\sqrt{2} \cr }
\right ) \otimes \left \{ \begin{array}{l} \left ( \matrix{ 1 & 0
& 0 \cr 0 & c & is \cr 0 & -is & c \cr } \right ) \; , \\ \\
\left ( \matrix{ c & 0 & is \cr 0 & 1 & 0 \cr -is & 0 & c \cr }
\right ) \; , \end{array} \right .
\end{equation}
from which $\theta^{}_{13} \neq 0^\circ$, $\theta^{}_{23} =
45^\circ$ and $\delta = 90^\circ$ can be predicted. Here
$c^{}_{12} \equiv \cos\theta^{}_{12}$, $s^{}_{12} \equiv
\sin\theta^{}_{12}$, $c \equiv \cos\theta$ and $s \equiv
\sin\theta$ are defined. The angle $\theta$ depends closely on the
parameters of $\mu$-$\tau$ symmetry breaking. An example for the
soft $\mu$-$\tau$ symmetry breaking will be given in section 5 of
my talk.

A generic parametrization of the $3\times 3$ Majorana neutrino
mixing matrix needs three mixing angles ($\theta^{}_{12}$,
$\theta^{}_{23}$, $\theta^{}_{13}$) and three CP-violating phases
($\delta$, $\rho$, $\sigma$) \cite{FX01}:
\begin{equation}
V = \left( \matrix{ c^{}_{12}c^{}_{13} & s^{}_{12}c ^{}_{13} &
s^{}_{13} e^{-i\delta} \cr -s^{}_{12}c^{}_{23}
-c^{}_{12}s^{}_{23}s^{}_{13} e^{i\delta} & c^{}_{12}c^{}_{23}
-s^{}_{12}s^{}_{23}s^{}_{13} e^{i\delta} & s^{}_{23}c^{}_{13} \cr
s^{}_{12}s^{}_{23} -c^{}_{12}c^{}_{23}s^{}_{13} e^{i\delta} &
-c^{}_{12}s^{}_{23} -s^{}_{12}c^{}_{23}s^{}_{13} e^{i\delta} &
c^{}_{23}c^{}_{13} \cr } \right) \left ( \matrix{ e^{i\rho } & 0 &
0 \cr 0 & e^{i\sigma} & 0 \cr 0 & 0 & 1 \cr } \right ) \; .
\end{equation}
One usually refers to $\delta$ as the ``Dirac" phase, because it
is also present in the $3\times 3$ Dirac neutrino mixing matrix.
In comparison, $\rho$ and $\sigma$ are commonly referred to as the
``Majorana" phases, which do not manifest themselves in neutrino
oscillations and can be rotated away if neutrinos are Dirac
particles. A complex $\mu$-$\tau$ symmetry breaking scheme is
possible to generate both non-trivial $\delta$ and non-trivial
$\rho$ and $\sigma$ in a given neutrino mass model.

This talk will focus on a new kind of flavor symmetry, the
so-called Friedberg-Lee (FL) symmetry \cite{Lee}, to discuss the
neutrino mass matrix and neutrino mixing. One will see that the
well-known $\mu$-$\tau$ symmetry can also manifest itself in the
neutrino mass operator with the FL symmetry.

\section{What is the FL symmetry?}

The effective neutrino mass operator proposed recently by
Friedberg and Lee \cite{Lee} is of the form
\begin{eqnarray}
{\cal L}^{}_{\rm FL} & = & a \left (\overline{\nu}^{}_\tau -
\overline{\nu}^{}_\mu \right ) \left (\nu^{}_\tau - \nu^{}_\mu
\right ) + b \left (\overline{\nu}^{}_\mu - \overline{\nu}^{}_e
\right ) \left (\nu^{}_\mu - \nu^{}_e \right ) + c \left
(\overline{\nu}^{}_e - \overline{\nu}^{}_\tau \right ) \left
(\nu^{}_e - \nu^{}_\tau \right ) ~~~~~~~
\nonumber \\
& & + ~ m^{}_0 \left (\overline{\nu}^{}_e \nu^{}_e +
\overline{\nu}^{}_\mu \nu^{}_\mu + \overline{\nu}^{}_\tau
\nu^{}_\tau \right ) \; ,
\end{eqnarray}
where $a$, $b$, $c$ and $m^{}_0$ are all assumed to be real. A
salient feature of ${\cal L}^{}_{\rm FL}$ is its partial
translational symmetry; i.e., its $a$, $b$ and $c$ terms are
invariant under the transformation $\nu^{}_\alpha \rightarrow
\nu^{}_\alpha + z$ (for $\alpha = e, \mu, \tau$) with $z$ being a
space-time independent constant element of the Grassmann algebra.
Corresponding to Eq. (9), the neutrino mass matrix $M^{}_\nu$
reads
\begin{equation}
M^{}_\nu \; = \; m^{}_0 \left ( \matrix{ 1 & ~0~ & 0 \cr 0 & 1 & 0
\cr 0 & 0 & 1 \cr } \right ) + \left ( \matrix{ b+c & -b & -c \cr
-b & ~ a+b ~ & -a \cr -c & -a & a+c \cr } \right ) \; .
\end{equation}
Diagonalizing $M^{}_\nu$ by the transformation $V^{\dagger}_{\rm
FL} M^{}_\nu V^{*}_{\rm FL} = {\rm Diag} \{ m^{}_1, m^{}_2, m^{}_3
\}$, in which $m^{}_i$ (for $i=1,2,3$) stand for the neutrino
masses, one may obtain the neutrino mixing matrix
\begin{equation}
V^{}_{\rm FL} = \left ( \matrix{ \frac{2}{\sqrt{6}} &
\frac{1}{\sqrt{3}} & 0 \cr -\frac{1}{\sqrt{6}} & ~
\frac{1}{\sqrt{3}} ~ & \frac{1}{\sqrt{2}} \cr -\frac{1}{\sqrt{6}}
& \frac{1}{\sqrt{3}} & -\frac{1}{\sqrt{2}} \cr } \right ) \left (
\matrix{ \cos\frac{\theta}{2} & 0 & \sin\frac{\theta}{2} \cr 0 & 1
& 0 \cr -\sin\frac{\theta}{2} & ~0~ & \cos\frac{\theta}{2} \cr }
\right ) \; ,
\end{equation}
where $\theta$ is given by $\tan\theta = \sqrt{3} \left (b-c
\right )/\left [ \left (b + c\right ) - 2a \right ]$. This
interesting result leads us to the following observations
\cite{XZZ}:

(1) If $\theta =0^\circ$ holds, $V^{}_{\rm FL}$ will reproduce the
exact tri-bimaximal neutrino mixing pattern \cite{TB}. The latter,
which can be understood as a geometric representation of the
neutrino mixing matrix \cite{TD}, is in good agreement with
current experimental data. Non-vanishing but small $\theta$
predicts $\sin\theta^{}_{13} = ( 2/\sqrt{6}) \sin \left (\theta/2
\right )$, implying $\theta \leq 24.6^\circ$ for $\theta^{}_{13} <
10^\circ$. On the other hand, $\theta^{}_{23}$ will mildly deviate
from its best-fit value $\theta^{}_{23} = 45^\circ$ if $\theta$
(or $\theta^{}_{13}$) takes non-zero values.

(2) The limit $\theta = 0^\circ$ results from $b = c$. When $b=c$
holds, it is straightforward to see that the neutrino mass
operator ${\cal L}^{}_{\rm FL}$ has the exact $\mu$-$\tau$
symmetry (i.e., ${\cal L}^{}_{\rm FL}$ is invariant under the
interchange of $\mu$ and $\tau$ indices). In other words, the
tri-bimaximal neutrino mixing is a natural consequence of the
$\mu$-$\tau$ symmetry of $M^{}_\nu$ in the FL model. Then
$\theta^{}_{13} \neq 0^\circ$ and $\theta^{}_{23} \neq 45^\circ$
measure the strength of $\mu$-$\tau$ symmetry breaking, as many
authors have discussed in other neutrino mass models.

Note that ${\cal L}^{}_{\rm FL}$ is only valid for Dirac
neutrinos. Recently Zhang, Zhou and I have generalized ${\cal
L}^{}_{\rm FL}$ to describe Majorana neutrinos \cite{XZZ}. In
particular, we allow $a$, $b$, $c$ and $m^{}_0$ to be complex so
as to accommodate leptonic CP violation. Two special scenarios
have been discussed in detail. In scenario (A), we require that
$a$ and $m^{}_0$ be real and $b = c^*$ be complex. We find that
the $\mu$-$\tau$ symmetry of $M^{}_\nu$ is softly broken in this
case, leading to the elegant predictions $\theta^{}_{13} \neq
0^\circ$, $\theta^{}_{23} =45^\circ$ and $\delta = 90^\circ$. Two
Majorana CP-violating phases $\rho$ and $\sigma$ keep vanishing.
In scenario (B), we assume that $a$, $b$ and $c$ are real but
$m^{}_0$ is complex. We find that the results of $\theta^{}_{12}$,
$\theta^{}_{23}$ and $\theta^{}_{13}$ obtained from $V^{}_{\rm
FL}$ keep unchanged in this case, but some non-trivial values of
the Majorana CP-violating phases $\rho$ and $\sigma$ can now be
generated. The Dirac CP-violating phase $\delta$ remains
vanishing.

To see the point of the FL symmetry more clearly, let me switch
off the term proportional to $m^{}_0$ and then rewrite ${\cal
L}^{}_{\rm FL}$ for Majorana neutrinos:
\begin{eqnarray}
{\cal L}^{}_{\rm mass} & = & \frac{1}{2} \left [ a
(\overline{\nu^{}_{\tau \rm L}}- \overline{\nu^{}_{\mu \rm L}})
(\nu^{\rm c}_{\tau \rm L} - \nu^{\rm c}_{\mu \rm L}) + b
(\overline{\nu^{}_{\mu \rm L}} - \overline{\nu^{}_{e \rm L}})
(\nu^{\rm c}_{\mu \rm L} - \nu^{\rm c}_{e \rm L}) \right .
\nonumber \\
&& \left. ~~~~~~~~~~~~~~~~~~~~~~~~~~~~~~~~~\;\;\; + c
(\overline{\nu^{}_{e \rm L}} - \overline{\nu^{}_{\tau \rm L}})
(\nu^{\rm c}_{e \rm L} - \nu^{\rm c}_{\tau \rm L}) \right ] + {\rm
h.c.} \; ,
\end{eqnarray}
where $a$, $b$ and $c$ are in general complex, and $\nu^{\rm
c}_{\alpha \rm L} \equiv C\overline{\nu^{}_{\alpha \rm L}}^T$ (for
$\alpha =e$, $\mu$, $\tau$). Corresponding to Eq. (12), the
Majorana neutrino mass matrix $M^{}_\nu$ is simply
\begin{equation}
M^{}_\nu = \left ( \matrix{ b+c & -b & -c \cr -b & ~ a+b ~ & -a
\cr -c & -a & a+c \cr } \right ) \; .
\end{equation}
The diagonalization of $M^{}_\nu$ is straightforward: $V^\dagger
M^{}_\nu V^* = \overline{M}^{}_\nu$, where $V$ is just the MNS
matrix, and $\overline{M}^{}_\nu = {\rm Diag} \{ m^{}_1, m^{}_2,
m^{}_3 \}$. From Eq. (13) together with the parametrization of $V$
in Eq. (8), it is easy to verify
\begin{equation}
{\rm Det}(M^{}_\nu) = {\rm Det} \left ( V \overline{M}^{}_\nu V^T
\right ) = {\rm Det} (\overline{M}^{}_\nu) \left [ {\rm Det}(V)
\right ]^2 = m^{}_1 m^{}_2 m^{}_3 e^{2 i \left (\rho + \sigma
\right )} = 0 \; .
\end{equation}
This result, which is an immediate consequence of the FL symmetry
in ${\cal L}^{}_{\rm mass}$, implies that one of the three
neutrinos must be massless. Because of $m^{}_2 > m^{}_1$ obtained
from the solar neutrino oscillation data \cite{SNO}, the $m^{}_2
=0$ case has been ruled out. In Eq. (9) the FL symmetry is broken
by $m^{}_0 \neq 0$, hence all the three neutrinos are massive. One
may wonder whether it is possible to break the FL symmetry but
keep $m^{}_1 =0$ or $m^{}_3 =0$ unchanged. The answer to this
question is actually affirmative, as one will see in the
subsequent section.

\section{Oblique FL symmetry breaking}

Is it really possible to break the FL symmetry in ${\cal
L}^{}_{\rm mass}$ but keep $m^{}_1 =0$ or $m^{}_3 =0$ unchanged?
Luo and I \cite{Luo} have found that the simplest way to make this
possibility realizable is to transform one of the neutrino fields
$\nu^{}_\alpha$ into $\kappa^* \nu^{}_\alpha$ with $\kappa \neq
1$. Given $\nu^{}_e \rightarrow \kappa^* \nu^{}_e$ as an example,
the resultant Majorana neutrino mass operator reads
\begin{eqnarray}
{\cal L}^\prime_{\rm mass} & = & \frac{1}{2} \left [ a
(\overline{\nu^{}_{\tau \rm L}}- \overline{\nu^{}_{\mu \rm L}})
(\nu^{\rm c}_{\tau \rm L} - \nu^{\rm c}_{\mu \rm L}) + b
(\overline{\nu^{}_{\mu \rm L}} - \kappa \overline{\nu^{}_{e \rm
L}}) (\nu^{\rm c}_{\mu \rm L} - \kappa \nu^{\rm c}_{e \rm L})
\right . \nonumber \\
&& \left . ~~~~~~~~~~~~~~~~~~~~~~~~~~~~~~~~~\;\;\; + c ( \kappa
\overline{\nu^{}_{e \rm L}} - \overline{\nu^{}_{\tau \rm L}}) (
\kappa \nu^{\rm c}_{e \rm L} - \nu^{\rm c}_{\tau \rm L}) \right ]
+ {\rm h.c.} \; .
\end{eqnarray}
Accordingly, the neutrino mass matrix is given by
\begin{equation}
M^\prime_\nu = \left ( \matrix{ \kappa^2 (b+c) & - \kappa b & -
\kappa c \cr - \kappa b & ~ a+b ~ & -a \cr - \kappa c & -a & a+c
\cr } \right ) \; .
\end{equation}
We are then left with ${\rm Det}(M^\prime_\nu) = \kappa^2 {\rm
Det}(M^{}_\nu) =0$, which is independent of the magnitude and
phase of $\kappa$. Thus we obtain either $m^{}_1 =0$ or $m^{}_3
=0$ from $M^\prime_\nu$. The effect of $\kappa \neq 1$ is referred
to as the ``oblique" symmetry breaking, because it breaks the FL
symmetry but keeps the lightest neutrino to be massless.

My next step is to show that a generic bi-large neutrino mixing
pattern can be derived from $M^\prime_\nu$. Let me focus on the
$m^{}_1 =0$ case,\footnote{The $m^{}_3 =0$ case is actually
disfavored, if one intends to achieve $\theta^{}_{23} = 45^\circ$
and $\theta^{}_{13} =0^\circ$ from $M^\prime_\nu$ in the
leading-order approximation.} in which $M^\prime_\nu$ is
diagonalized by the transformation $V^\dagger M^\prime_\nu V^* =
\overline{M}^{}_\nu$ with $\overline{M}^{}_\nu = {\rm Diag}\{0,
m^{}_2, m^{}_3 \}$. As the best-fit values of the atmospheric and
CHOOZ neutrino mixing angles are $\theta^{}_{23} = 45^\circ$ and
$\theta^{}_{13} =0^\circ$ respectively \cite{Vissani}, we may
decompose $V$ into a product of two special unitary matrices: $V =
U O$, where
\begin{equation}
U = \left ( \matrix{ \frac{1}{\sqrt{2|\kappa|^2 +1}} &
\frac{\sqrt{2} \; \kappa}{\sqrt{2|\kappa|^2 +1}} & ~~~ 0 \cr
\frac{\kappa^*}{\sqrt{2|\kappa|^2 +1}} & ~~~ -\frac{1}{\sqrt{2
\left ( 2|\kappa|^2 +1 \right )}} ~ & ~\;\; \frac{1}{\sqrt{2}} \cr
\frac{\kappa^*}{\sqrt{2|\kappa|^2 +1}} & ~~ -\frac{1}{\sqrt{2
\left (2|\kappa|^2 +1 \right )}} ~ & ~ -\frac{1}{\sqrt{2}} \cr }
\right ) \; ,
\end{equation}
and $O$ depends on the details of $M^\prime_\nu$. Note that the
first column of $U$ is associated with $m^{}_1 =0$. $O$ contains a
single $(2,3)$-rotation angle, which can be obtained from the
diagonalization of $U^\dagger M^\prime_\nu U^*$.

For simplicity, Luo and I have assumed that $a$, $b$ and $c$ are
all real \cite{Luo}. In this case, $U^\dagger M^\prime_\nu U^*$ is
a real symmetric matrix and the calculation of $O$ is quite easy.
We arrive at
\begin{equation}
V = \left ( \matrix{ \frac{1}{\sqrt{2|\kappa|^2 +1}} &
\frac{\sqrt{2} \; \kappa \cos\theta}{\sqrt{2|\kappa|^2 +1}} &
\frac{\sqrt{2} \; \kappa \sin\theta}{\sqrt{2|\kappa|^2 +1}} \cr
\frac{\kappa^*}{\sqrt{2|\kappa|^2 +1}} & ~ -\frac{1}{\sqrt{2}}
\left ( \frac{\cos\theta}{\sqrt{2|\kappa|^2 +1}} + \sin\theta
\right ) ~ & ~\;\; \frac{1}{\sqrt{2}} \left ( \cos\theta -
\frac{\sin\theta}{\sqrt{2|\kappa|^2 +1}} \right ) \cr
\frac{\kappa^*}{\sqrt{2|\kappa|^2 +1}} & -\frac{1}{\sqrt{2}} \left
( \frac{\cos\theta}{\sqrt{2|\kappa|^2 +1}} - \sin\theta \right ) &
-\frac{1}{\sqrt{2}} \left ( \cos\theta
+\frac{\sin\theta}{\sqrt{2|\kappa|^2 +1}} \right ) \cr } \right )
\; ,
\end{equation}
where $\theta$ is given by $\tan 2\theta = \left (b-c \right )
\sqrt{ 2|\kappa|^2 + 1}/\left [ \left (b+c \right ) |\kappa|^2 - 2
a \right ]$. One can see that $b =c$, which is a clear reflection
of the $\mu$-$\tau$ permutation symmetry in ${\cal L}^\prime_{\rm
mass}$ or $M^\prime_\nu$, leads to $\theta =0^\circ$ or
equivalently $\theta^{}_{23} = 45^\circ$ and $\theta^{}_{13}
=0^\circ$. After rephasing the expression of $V$ with the
transformations of charged-lepton and neutrino fields $e
\rightarrow e^{i\phi^{}_\kappa} e$, $\mu \rightarrow -\mu$,
$\nu^{}_1 \rightarrow e^{i\phi^{}_\kappa} \nu^{}_1$ and $\nu^{}_3
\rightarrow -\nu^{}_3$, where $\phi^{}_\kappa \equiv \arg
(\kappa)$, we may directly compare it with Eq. (8). A novel
prediction turns out to be \cite{Luo}
\begin{eqnarray}
\sin\theta^{}_{13} & = & \left |\frac{1 - \tan\theta^{}_{23}}{1+
\tan\theta^{}_{23}} \right | \tan\theta^{}_{12} \; .
\end{eqnarray}
Since the mixing angles $\theta^{}_{12}$ and $\theta^{}_{23}$ are
already known to a reasonable degree of accuracy, they can be used
to determine the unknown mixing angle $\theta^{}_{13}$. Eq. (19)
indicates that the deviation of $\theta^{}_{13}$ from zero is
closely correlated with the deviation of $\theta^{}_{23}$ from
$45^\circ$. Such a correlation can easily be tested in the near
future. On the other hand, Eq. (18) allows us to achieve
\begin{eqnarray}
|\kappa| \; = \; \frac{\sin\theta^{}_{12}}{\sqrt{\cos
2\theta^{}_{12} + \sin 2\theta^{}_{23}}} \;\; .
\end{eqnarray}
By using the $99\%$ confidence-level ranges of $\theta^{}_{12}$
and $\theta^{}_{23}$ (i.e., $30^\circ \leq \theta^{}_{12} \leq
38^\circ$ and $36^\circ \leq \theta^{}_{23} \leq 54^\circ$)
\cite{Vissani}, we obtain $\theta^{}_{13} \leq 7.1^\circ$ from Eq.
(19) and $0.41 \leq |\kappa| \leq 0.56$ from Eq. (20). Note that
$|\kappa| =1/2$ is in particular favorable and it implies that $U$
takes the tri-bimaximal mixing pattern \cite{TB}. The numerical
dependence of $\theta^{}_{13}$ on $\theta^{}_{12}$ and
$\theta^{}_{23}$ is illustrated in Fig. 1. The upper bound on
$\theta^{}_{13}$ extracted from Fig. 1 is certainly more stringent
than $\theta^{}_{13} < 10^\circ$ obtained from a global analysis
of current neutrino oscillation data \cite{Vissani}. It will be
interesting to see whether our prediction for the correlation
between the unknown mixing angle $\theta^{}_{13}$ and two known
angles can survive the future measurements.
\begin{figure}[t]
\begin{center}
\includegraphics[width=9cm,height=9cm]{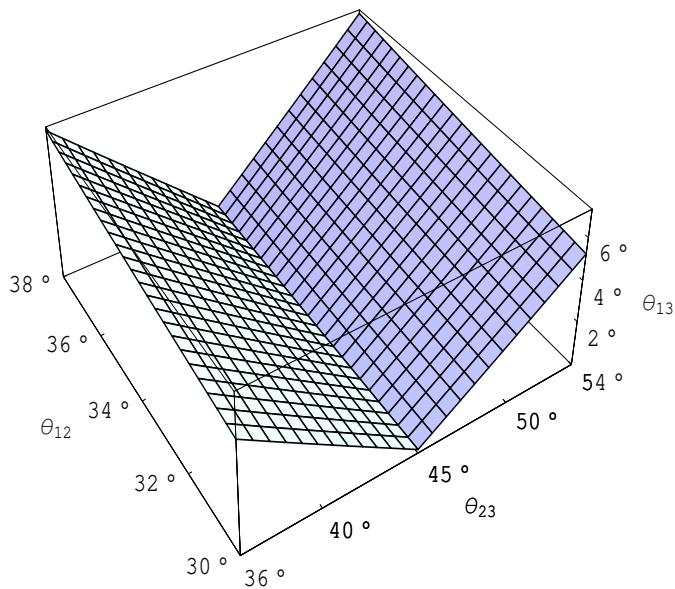}
\vspace{-0.5cm} \caption{Numerical dependence of $\theta^{}_{13}$
on $\theta^{}_{12}$ and $\theta^{}_{23}$ as predicted by Eq.
(19).}
\end{center}
\end{figure}

We have taken $\nu^{}_e \rightarrow \kappa^* \nu^{}_e$ to break
the FL symmetry and achieve a realistic pattern of the neutrino
mass matrix with $m^{}_1 =0$. A careful analysis shows that
neither $\nu^{}_\mu \rightarrow \kappa^* \nu^{}_\mu$ nor
$\nu^{}_\tau \rightarrow \kappa^* \nu^{}_\tau$ with $\kappa \neq
1$, which automatically breaks the $\mu$-$\tau$ permutation
symmetry, is favored to reproduce the exactly or approximately
tri-bimaximal neutrino mixing pattern.

\section{CP violation and seesaw realization}

Although the above discussions are based on the assumption of real
$a$, $b$ and $c$, they can easily be extended to the case of
complex $a$, $b$ and $c$ in order to accommodate CP violation. For
simplicity of illustration, here we assume that $a$ remains real
but $b = c^*$ is complex. We may then simplify the expression of
$U^\dagger M^\prime_\nu U^*$ by taking into account $b+c = 2{\rm
Re}(b)$ and $b-c = 2i {\rm Im}(b)$. After an analogous
calculation, we obtain the MNS matrix \cite{Luo}
\begin{equation}
V = \left ( \matrix{ \frac{1}{\sqrt{2|\kappa|^2 +1}} & i
\frac{\sqrt{2} \; \kappa \cos\theta}{\sqrt{2|\kappa|^2 +1}} & i
\frac{\sqrt{2} \; \kappa \sin\theta}{\sqrt{2|\kappa|^2 +1}} \cr
\frac{\kappa^*}{\sqrt{2|\kappa|^2 +1}} & ~ -\frac{1}{\sqrt{2}}
\left ( i \frac{\cos\theta}{\sqrt{2|\kappa|^2 +1}} + \sin\theta
\right ) ~ & ~\;\; \frac{1}{\sqrt{2}} \left ( \cos\theta - i
\frac{\sin\theta}{\sqrt{2|\kappa|^2 +1}} \right ) \cr
\frac{\kappa^*}{\sqrt{2|\kappa|^2 +1}} & -\frac{1}{\sqrt{2}} \left
( i \frac{\cos\theta}{\sqrt{2|\kappa|^2 +1}} - \sin\theta \right )
& -\frac{1}{\sqrt{2}} \left ( \cos\theta + i
\frac{\sin\theta}{\sqrt{2|\kappa|^2 +1}} \right ) \cr } \right )
\; ,
\end{equation}
where $\theta$ is given by $\tan 2\theta = - {\rm Im}(b) \sqrt{
2|\kappa|^2 + 1}/\left [a + {\rm Re}(b)\left (|\kappa|^2 + 1
\right ) \right ]$. Two immediate but important observations are
in order:

(1) In this simple scenario $V$ contains two nontrivial
CP-violating phases: $\delta = - 90^\circ$ (when $\theta
<0^\circ$) or $\delta = +90^\circ$ (when $\theta
>0^\circ$) and $\sigma = - 90^\circ$. Both of them are attributed to the purely
imaginary term $b-c$. The Jarlskog rephasing invariant of leptonic
CP violation\cite{J} reads ${\cal J} = |\kappa|^2 |\sin
2\theta|/\left [2 \left (2|\kappa|^2 +1 \right )^{3/2} \right ]$.
A numerical analysis yields $0.41 \leq |\kappa| \leq 0.57$ and
$|\theta| < 19.4^{\circ}$. Thus we arrive at ${\cal J} \leq
0.041$.

(2) $\tan \theta^{}_{23} = 1$ or $\theta^{}_{23} = 45^\circ$ can
be achieved, although the effective Majorana neutrino mass
operator ${\cal L}^\prime_{\rm mass}$ does not possess the exact
$\mu$-$\tau$ permutation symmetry. The reason is simply that $|b|
= |c|$ holds in our scenario. In other words, the phase difference
between $b$ and $c$ signifies a kind of soft $\mu$-$\tau$ symmetry
breaking which can keep $\theta^{}_{23} = 45^\circ$ but cause
$\theta^{}_{13} \neq 0^\circ$.

Finally I like to point out that it is possible to derive the
Majorana neutrino mass operator ${\cal L}^\prime_{\rm mass}$ from
the minimal seesaw model (MSM) \cite{MSM}, a canonical extension
of the standard model with only two heavy right-handed Majorana
neutrinos. The neutrino mass term in the MSM can be written as
\begin{equation}
-{\cal L}^{}_{\rm MSM} \; = \; \frac{1}{2} \overline{\left
(\nu^{~}_{\rm L}, ~N^{\rm c}_{\rm R} \right )} \left ( \matrix{
{\bf 0} & M^{~}_{\rm D} \cr M^T_{\rm D} & M^{~}_{\rm R} \cr }
\right ) \left ( \matrix{ \nu^{\rm c}_{\rm L} \cr N^{}_{\rm R} \cr
} \right ) ~ + ~ {\rm h.c.} \; ,
\end{equation}
where $\nu^{~}_{\rm L}$ and $N^{}_{\rm R}$ denote the column
vectors of $(\nu^{~}_e, \nu^{~}_\mu, \nu^{~}_\tau)^{~}_{\rm L}$
and $(N^{~}_1, N^{~}_2)^{}_{\rm R}$ fields, respectively. Provided
the mass scale of $M^{}_{\rm R}$ is considerably higher than that
of $M^{}_{\rm D}$, one may obtain the effective (left-handed)
Majorana neutrino mass matrix $M^\prime_\nu$ from Eq. (22) via the
well-known seesaw mechanism \cite{SS}: $M^\prime_\nu = M^{}_{\rm
D} M^{-1}_{\rm R} M^T_{\rm D}$. As $M^{}_{\rm R}$ is of rank 2,
${\rm Det}(M^\prime_\nu) =0$ holds and $m^{}_1 =0$ (or $m^{}_3
=0$) is guaranteed. We find that the expression of $M^\prime_\nu$
given in Eq. (16) can be reproduced from $M^{}_{\rm D}$ and
$M^{}_{\rm R}$ if they take the following forms:
\begin{eqnarray}
M^{}_{\rm D} & = & \Lambda^{}_{\rm D} \left ( \matrix{ \kappa & 0
\cr -1 & -1 \cr 0 & 1 \cr } \right ) \; , \nonumber \\
M^{}_{\rm R} & = & \frac{\Lambda^2_{\rm D}}{ab + bc + ca} \left (
\matrix{ ~ a+c ~ & c \cr c & ~ b+c ~ \cr } \right ) \; ,
\end{eqnarray}
where $\Lambda^{}_{\rm D}$ characterizes the mass scale of
$M^{}_{\rm D}$. Such textures are by no means unique, but they may
serve as a good example to illustrate how the seesaw mechanism
works to give rise to $M^\prime_\nu$ or ${\cal L}^\prime_{\rm
mass}$ in the MSM.

\section{Concluding remarks}

I have argued that the observed flavor structures of leptons and
quarks might imply the existence of certain flavor symmetries. A
flavor symmetry itself is far away from a flavor theory, but it
might be taken as a reasonable starting point to build a realistic
flavor model towards deeper understanding of the fermion mass
spectra and flavor mixing patterns. The $\mu$-$\tau$ permutation
symmetry has been taken as an example to interpret the almost
maximal atmospheric neutrino mixing angle ($\theta^{}_{23} \sim
45^\circ$) and the strongly suppressed CHOOZ neutrino mixing angle
($\theta^{}_{13} < 10^\circ$). The effect of $\mu$-$\tau$ symmetry
breaking could even show up in the ultrahigh-energy neutrino
telescopes \cite{TELE}.

Let us anticipate that IceCube \cite{Ice} and other
second-generation neutrino telescopes \cite{Water} are able to
detect the fluxes of ultrahigh-energy (UHE) $\nu^{}_e$
($\overline{\nu}^{}_e$), $\nu^{}_\mu$ ($\overline{\nu}^{}_\mu$)
and $\nu^{}_\tau$ ($\overline{\nu}^{}_\tau$) neutrinos generated
from very distant astrophysical sources. For most of the
currently-envisaged sources of UHE neutrinos \cite{R}, a general
and canonical expectation is that the initial neutrino fluxes are
produced via the decay of pions created from $pp$ or $p\gamma$
collisions and their flavor content can be expressed as
\begin{equation}
\left \{\phi^{}_e ~,~ \phi^{}_\mu ~,~ \phi^{}_\tau \right \} \; =
\; \left \{ \frac{1}{3} ~,~ \frac{2}{3} ~,~ 0 \right \} \phi^{}_0
\; ,
\end{equation}
where $\phi^{}_\alpha$ (for $\alpha = e, \mu, \tau$) denotes the
sum of $\nu^{}_\alpha$ and $\overline{\nu}^{}_\alpha$ fluxes, and
$\phi^{}_0 = \phi^{}_e + \phi^{}_\mu + \phi^{}_\tau$ is the total
flux of neutrinos and antineutrinos of all flavors. Due to
neutrino oscillations, the flavor composition of such cosmic
neutrino fluxes to be measured at the detector of a neutrino
telescope has been expected to be \cite{Pakvasa}
\begin{equation}
\left \{\phi^{\rm D}_e ,~ \phi^{\rm D}_\mu ,~ \phi^{\rm D}_\tau
\right \} \; = \; \left \{ \frac{1}{3} ~,~ \frac{1}{3} ~,~
\frac{1}{3} \right \} \phi^{}_0 \; .
\end{equation}
However, it is worth remarking that this naive expectation is only
true in the limit of $\mu$-$\tau$ symmetry (or equivalently,
$\theta^{}_{13} = 0$ and $\theta^{}_{23} = \pi/4$). Starting from
the hypothesis given in Eq. (24) and allowing for the slight
breaking of $\mu$-$\tau$ symmetry, I have shown that \cite{TELE}
\begin{eqnarray}
\phi^{\rm D}_e : \phi^{\rm D}_\mu : \phi^{\rm D}_\tau  = \left (1
-2 \Delta \right ) : \left (1 +\Delta \right ) : \left (1 +\Delta
\right ) \;
\end{eqnarray}
holds to an excellent degree of accuracy, where
\begin{equation}
\Delta \; = \; \frac{1}{4} \left ( 2\varepsilon \sin^2
2\theta^{}_{12} - \theta^{}_{13} \sin 4\theta^{}_{12} \cos\delta
\right ) \;
\end{equation}
with $\varepsilon \equiv \theta^{}_{23} - \pi/4$ characterizes the
effect of $\mu$-$\tau$ symmetry breaking (i.e., the combined
effect of $\theta^{}_{13} \neq 0$ and $\theta^{}_{23} \neq
\pi/4$). I obtain $-0.1 \leq \Delta \leq +0.1$ from the present
neutrino oscillation data \cite{Vissani}. It would be fantastic if
$\Delta$ could finally be measured at a $\rm km^2$-scale neutrino
telescope in the future.

The main body of this talk is actually to highlight the FL
symmetry and its oblique breaking scheme. I have emphasized that
the FL symmetry is a new kind of flavor symmetry applicable to the
building of neutrino mass models. Imposing this symmetry on the
effective Majorana neutrino mass operator, Luo and I have shown
that it can be broken in such a novel way that the lightest
neutrino remains massless but an experimentally-favored bi-large
neutrino mixing pattern is achievable. This phenomenological
scenario predicts a testable relationship between the unknown
neutrino mixing angle $\theta^{}_{13}$ and the known angles
$\theta^{}_{12}$ and $\theta^{}_{23}$ in the CP-conserving case:
$\sin\theta^{}_{13} = \tan\theta^{}_{12} |(1- \tan\theta^{}_{23})/
(1+ \tan\theta^{}_{23})|$. Such a result is suggestive and
interesting because it directly correlates the deviation of
$\theta^{}_{13}$ from zero with the deviation of $\theta^{}_{23}$
from $\pi/4$. We have discussed a simple but instructive
possibility of introducing CP violation into the Majorana neutrino
mass operator, in which the soft breaking of $\mu$-$\tau$ symmetry
yields $\delta = \pi/2$ (or $\delta = -\pi/2$) but keeps
$\theta^{}_{23} =\pi/4$. We have also discussed the possibility of
incorporating our scenario in the MSM.

In conclusion, the FL symmetry and its breaking mechanism may have
a wealth of implications in neutrino phenomenology. The physics
behind this new flavor symmetry remains unclear and deserves a
further study.

\section*{Acknowledgements}

I would like to thank Y. Koide for inviting me to this stimulating
workshop in Shizuoka. I am also grateful to S. Luo, H. Zhang and
S. Zhou for their collaboration in the study of the Friedberg-Lee
symmetry. This work was supported in part by the National Natural
Science Foundation of China.

\end{document}